\documentclass[square]{ws-procs11x85}
\usepackage{balance} 

\def\Journal#1#2#3#4{{#1} {\bf #2}, #3 (#4)}

\begin{document}
\twocolumn[
\title{Particle acceleration close to the supermassive black hole horizon: the case of M87}

\author{F.M. Rieger}
\address{Max-Planck-Institut f\"ur Kernphysik, Saupfercheckweg 1, 69117 Heidelberg, Germany;\\
and European Associated Laboratory for Gamma-Ray Astronomy, jointly supported by CNRS and MPG,\\
E-mail: frank.rieger@mpi-hd.mpg.de}

\author{F.A. Aharonian}

\address{Dublin Institute of Advanced Studies, 31 Fitzwilliam Place, Dublin 2, Ireland}
\begin{abstract}
The radio galaxy M87 has recently been found to be a rapidly variable TeV emitting source. 
We analyze the implications of the observed TeV characteristics and show that it proves 
challenging to account for them within conventional acceleration and emission models. 
We discuss a new pulsar-type scenario for the origin of variable, very high energy (VHE) 
emission close to the central supermassive black hole and show that magneto-centrifugally 
accelerated electrons could efficiently Compton upscatter sub-mm ADAF disk photons to the 
TeV regime, leading to VHE characteristics close to the observed ones. This suggests, 
conversely, that VHE observations of highly under-luminous AGNs could provide an important 
diagnostic tool for probing the conditions prevalent in the inner accretion disk of these 
sources.
\end{abstract}
\vskip12pt  
]
\bodymatter
\section{Introduction}
The advance in modern imaging atmospheric Cherenkov technologies has opened up a new
window on the non-thermal Universe, providing a unique diagnostic tool to study the 
physical conditions in violent regions of galactic and extragalactic sources. Until
recently, blazars have been the only extragalactic sources known to emit TeV ($10^{12}$
eV) radiation. Although comprising only a small fraction of radio-loud active galactic 
nuclei (AGNs), the favourable orientation of their jets at small viewing angles $i$ to 
the line of sight ($i < 10^{\circ}$), combined with the bulk relativistic motion of 
the plasma in their jets, makes them a privileged source class on the TeV sky. The 
H.E.S.S. detection of variable TeV emission from the non-blazar ($i \gtrsim 19^{\circ}$) 
radio galaxy M87 (redshift $z=0.0043$, distance $d \sim 16$ Mpc) during a campaign in
2003-2006 \cite{aha06}, may thus come surprising. Although some detectable TeV emission 
has been anticipated based on simple homogeneous SSC model extrapolations \cite{bai01}, 
the observed rapid $\gamma$-ray flux variations and the hard TeV spectrum comes quite 
unexpected. The giant elliptical galaxy M87 is known to host one of the most massive 
black holes $M_{\rm BH} \simeq 3 \times 10^9\,M_{\odot}$ in the Universe, and a prominent, 
non-aligned jet detectable from radio to X-ray wavelengths. As we show below, the VHE 
results may point to a new acceleration process occurring in the vicinity of the central 
supermassive black hole (BH).  
\section{Observed VHE characteristics in M87 and implications}
The H.E.S.S. observations of M87 in 2005 \cite{aha06} show a gamma-ray spectrum that 
reaches beyond 10 TeV and is consistent with a relatively hard power-law (with spectral 
index $\alpha \sim 1.2$, where $S_{\nu} \propto \nu^{-\alpha}$). The detected VHE 
output is quite moderate, with an isotropic TeV luminosity of $\sim 3 \times 10^{40}$ 
erg/s. During the 2005 high state, significant variability (flux doubling) on time scales 
of $\Delta t_{\rm obs} \sim 2$ days was found, the fastest variability observed in any 
waveband from M87 so far.\\ 
Several scenarios for the origin of VHE emission from M87 have been discussed in the 
literature, ranging from the vicinity of the central supermassive BH \cite{ner07,rie08} 
and the innermost part of the jet \cite{geo05,len08,tav08} to the superluminal feature 
HST-1 at a de-projected distance of $\sim 100$ pc from the central BH \cite{sta06,che07}. 
The latter has some interesting features including the fact that the TeV flux 
maximum in 2005 seems to coincide with a Chandra X-ray flux maximum from HST-1. 
On the other hand, the same observations \cite{har06,aha06} also indicate that the 
TeV and HST-1 X-ray flux evolutions from 2003 to 2004 are not correlated, yet the 
TeV spectral shape does not change (i.e., same power index within errors) from 
2004 to 2005, so that an HST-1 origin of the TeV emission in 2005 seems to require
some sort of cosmic conspiracy. Note further, that the observed spectrum of HST-1 
appears synchrotron cooling dominated above the break at $\nu_{\rm b} \simeq 10^{15}$ 
Hz, where the spectrum steepens to $\alpha \simeq 1.0$ ($S_{\nu} \propto \nu^{-\alpha}$) 
\cite{har06}, corresponding to an average radiating particle distribution $n(\gamma') 
\propto \gamma'^{-3}$. In order to Compton upscatter lower energy photons to the 
observed TeV regime, the presence of energetic electrons with Lorentz factors up to 
$\gamma_e' \sim 10^7/\delta$ is required within the source region, where $\delta$ is 
the bulk Lorentz factor of the source. If scattering in a single SSC-type scenario 
occurs in the Thomson limit, the target photon field, and thus the observed TeV flux, 
is expected to vary on time scales larger than implied by a source limit inferred 
from the cooling break, i.e., the total source size would be at most comparable to
$R \sim c\,\Delta t_{\rm obs}\,\delta$. If, on the other hand, the relevant scattering 
occurs in the Klein-Nishima regime, the resultant TeV spectrum would be very steep, 
contrary to the observed one. Further studies are required to assess whether current 
HST-1 models are flexible enough to overcome these issues.\\ 
A similar consideration seems to hold for leptonic scenarios that assume Compton 
scattering in a decelerating relativistic inner jet flow \cite{geo05}. Radio
features in the inner jet, for example, seem to be at most mildly relativistic
\cite{ly07,kov07}, yet significantly relativistic on larger scale \cite{che07}, 
the jet thus more resembling an accelerating than a decelerating flow, although 
matters may be more complicated if an internal velocity structure (e.g., spine 
plus sheath) is present. Note also, that the predicted spectral shapes for the 
TeV regime in simple model applications are much steeper than observed, although 
again, an internal jet structure may help by introducing additional acceleration 
and radiation effects \cite{rie04,tav08}.\\ 
Here we consider an alternative scenario assuming particle acceleration and TeV 
$\gamma$-ray production to take place close to the event horizon of the central 
supermassive black hole. We show that such a model can successfully reproduce 
the observed VHE characteristics and introduce an important link between accretion 
disk physics and jet formation theory. 
\section{Particle acceleration close to the central black hole}
Today, magneto-hydrodynamical models are widely considered to represent the most 
promising class for the formation and collimation of relativistic astrophysical 
jets. According to this picture, magnetic flux dragged inward and amplified by 
dynamo actions in the inner accretion disk can build up a rigidly rotating, dipolar 
magnetosphere. Along open flux surfaces the bulk of the plasma is centrifugally 
accelerated to relativistic speeds ($\Gamma_b \sim 10$) and collimated outside the 
light cylinder $r_{\rm L} \simeq (5-10)\, r_s$, e.g., see \cite{fen97,cam99}. It 
has been realized for quite a while \cite{gan97,rie00,osm07}, that such a MHD 
field structure could also allow for efficient centrifugal acceleration of test 
particles, i.e., a test particle co-rotating with the field line (bead-on-wire 
motion) will experience the centrifugal force and gain rotational energy while 
moving outwards \cite{mac94,che96}. The radial motion of such a particle is most 
conveniently analyzed in the framework of Hamiltonian dynamics \cite{rie00,rie08}. 
Consider an idealized two-dimensional model topology: since the Lagrangian $L$ for 
a particle with rest mass $m_0$ on a relativistically, rigidly rotating wire 
(angular velocity $\Omega=c/r_{\rm L}=$ constant) is not explicitly time-dependent, 
i.e.,
\begin{equation}
L=-m_0 c^2 \sqrt{1-r^2/r_{\rm L}^2-\dot{r}^2/c^2}\,,
\end{equation} the associated Hamiltonian 
\begin{equation}\label{hamiltonian}
H=\gamma\,m_0\,c^2 (1-r^2/r_{\rm L}^2)\,= \mathrm{constant}\,,
\end{equation} with $\gamma=(1-r^2/r_{\rm L}^2-\dot{r}^2/c^2)^{-1/2}$ the Lorentz factor, 
is a constant of motion. Thus, as a particle approaches $r_{\rm L}$, the term in 
brackets of Eq.~(\ref{hamiltonian}) gets smaller and has to be compensated by an 
increase in $\gamma$, which implies that the Lorentz factor increases dramatically 
for a particles approaching the light cylinder. In reality, unlimited growth will 
be prohibited by radiative energy losses (e.g., inverse Compton in the ambient 
photon field), the breakdown of the bead-on-the-wire (BW) approximation or the 
bending of the field line with increasing inertia \cite{gan97,rie00}. 
In particular, validity of the BW approximation requires, that the characteristic 
acceleration time scale, which can be easily derived from Eq.~(\ref{hamiltonian}),
i.e.,
\begin{equation}
t_{\rm acc}=\frac{\gamma}{\dot{\gamma}} 
  \simeq \frac{1}{2\,\Omega\,\tilde{m}^{1/4}\gamma^{1/2}}\,,
\end{equation} is always larger than the inverse of the relativistic gyro-frequency 
$\omega=e\,B/\gamma\,m_0\,c$. Here $\tilde{m}$ is determined by the initial
conditions, i.e. $\tilde{m}=1/(\gamma_0^2\,[1-r_0^2/r_{\rm L}^2]^2)$. 
This constrains achievable Lorentz factors to 
\begin{equation}
 \gamma_{\rm max}^{\rm BB} \leq \frac{1}{\tilde{m}^{1/6}} 
        \left(\frac{q\,B}{2\,m_0\,c^2}\,r_{\rm L}\right)^{2/3}\,
\end{equation} which is formally equivalent to the requirement that the Coriolis force 
must not exceed the Lorentz force \cite{rie00}. For parameter relevant to 
M87 (i.e., $B(r_{\rm L}) \sim 10 $ G, $r_{\rm L} \sim 5 \times 10^{15}$ cm), 
one finds $\gamma_{\rm max}^{\rm BB} \sim 5 \times 10^8$ for electrons. 
In general, centrifugal acceleration is less favourable for protons in the 
sense that $\gamma_{\rm max}^{\rm BB} \propto m_0^{-2/3}$ indicating that 
accelerated protons will not be able to interact efficiently with the ambient 
photon field.
\section{Inverse Compton upscattering of Comptonized disk photons}
\begin{figure}[t]
\center
\centerline{\psfig{figure=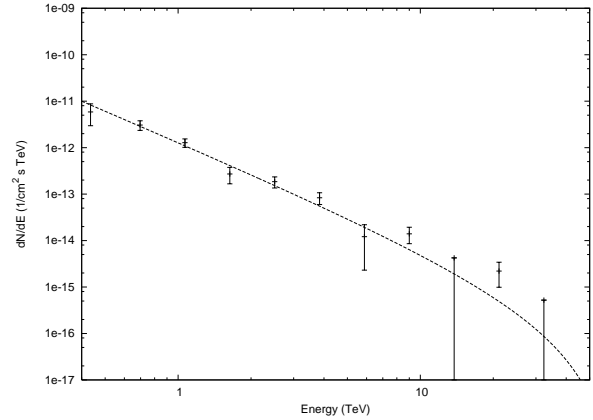,width=8truecm}}
\caption{Differential energy spectrum of M87 for the H.E.S.S. 2005 data. Error 
bars given are for statistical errors only. Shown is a characteristic TeV 
spectrum in the range $4\gamma_2^2  \nu_1 '  \leq \nu \leq 4\gamma_2^2 \nu_2'$, 
resulting from (Thomson) Compton upscattering of a seed photon spectrum with 
spectral index $\alpha_c=1.2$ between $\nu_1' \leq \nu' \leq \nu_2'$, with 
$\nu_1'/\nu_2'=0.015$, $p=2$ and $\gamma_2 \sim 2 \times 10^7$.}\label{fig1}
\end{figure}

In realistic astrophysical environments, radiative energy losses will always 
compete with acceleration and introduce additional constraints on the maximum
achievable particle energies. Given its very low bolometric luminosity output, 
mass accretion in M87 is likely to occur in a two-temperature, advection-dominated 
(ADAF) mode \cite{rey96,cam99,dim03}. For the estimated Bondi accretion rate 
of $\dot{m} \simeq 1.6 \times 10^{-3} \dot{m}_{\rm Edd}$ \cite{dim03}, the 
characteristic ADAF spectrum close to $r_{\rm L}$ peaks at around $\nu_s' \sim
10^{11}$ Hz with an associated (radio) luminosity $L_{\rm R} \sim 10^{40}$ erg/s. 
Inverse Compton interactions with the ambient photon field are thus not expected
to lead to a much stronger constraint on the electron Lorentz factors compared 
with $\gamma_{\rm max}^{\rm BB}$, suggesting that electron Lorentz factors up to
$\gamma \sim (10^7-10^8)$ could well be achieved via the proposed acceleration
mechanism \cite{rie08}. Hence, Compton (Thomson) upscattering of sub-mm ($\nu'
\leq \nu_2' \sim 10^{13}$ Hz) accretion disk photons can easily result in VHE 
photons with energies extending up to $\sim 10$ TeV and beyond. 
Note that this is different from the case of luminous AGN sources, where severe 
inverse Compton losses tend to limit achievable electron Lorentz factors to be 
below one thousand \cite{gan97,rie00}. Yet, even for the latter sources, 
centrifugal acceleration could operate as a promising pre-acceleration 
mechanism, possibly along with electrostatic wave surfing \cite{die06}, and
thus provide the energetic seed particles required for efficient Fermi 
acceleration on larger scales.\\
For the inferred Bondi accretion rate $\dot{m}_B$ in M87, comptonization of 
cyclosynchrotron soft photons in an ADAF is expected to add a power law-like 
tail to the disk spectrum above $\nu_s'$ with spectral index close to $\alpha_c 
\sim 1.2$ \cite{mah97,rie08}. Suppose that energetic electrons, after being 
released from the acceleration process with high Lorentz factors, encounter 
such a comptonized disk photon field. In general, the shape of the emergent (singly, 
Thomson) scattered inverse Compton spectrum $j_{\rm IC}(\nu)$ will depend on 
both the seed photon spectrum $F(\nu')$ and the differential electron 
distribution $n(\gamma)$, e.g.,
\begin{eqnarray}
j_{\rm IC}(\nu) & \propto & \nu^{-(p-1)/2} \int_{\nu_1'}^{\nu_2'} d\nu'
                   F(\nu')\,\nu'^{(p-1)/2} \nonumber\\
                & \times& \int_{\rm min\{x_2,1\}}^{\rm min\{x_1,1\}}dx\,
                   x^{(p-1)/2} f_{\rm IC}(x)\,,
\end{eqnarray} assuming a power law electron distribution $n(\gamma) \propto \gamma^{-p}$ 
in the interval $\gamma_{1} \leq \gamma \leq \gamma_{2} \sim 5 \times 10^7$, 
with $x_i:=\nu/(4\gamma_i^2\nu')$ and $f_{\rm IC}(x)=2x\ln x+x+1-2x^2$. Far away 
from the endpoints, i.e., for $\nu \ll 4\gamma_2^2 \nu_s' \sim 5$ TeV, we obtain 
the common $j_{\rm IC}(\nu) \propto \nu^{-(p-1)/2}$ power law evolution, 
independent of the incident photon spectrum. Yet, closer to the endpoints, i.e., 
for $4 \gamma_2^2 \nu_s' < \nu < 4 \gamma_2^2 \nu_2'$, we become again sensitive 
to the seed photon distribution so that the resultant inverse Compton VHE spectrum 
will be power law-like with spectral index $\alpha \simeq \alpha_c$. This is 
illustrated in Fig.~\ref{fig1} for $p=2$.
\section{Discussion}
Taking the detected TeV emission in M87 to arise via inverse Compton scattering 
by centrifugally accelerated electrons close to the central black hole, requires 
to examine whether, amongst others, the model can self-consistently account for 
the observed variability and luminosity output:\\
(1) The observed flux doubling time scale is of order $\sim 2$ days. Centrifugal
acceleration, on the other hand, occurs on a characteristic time scale $r_{\rm L}/c$. 
The observed variability thus implies a light cylinder radius of order $r_{\rm L} 
\sim 6\,r_s$, which is indeed consistent with expectations from relativistic MHD 
jet formation models \cite{fen97,cam99}.\\
(2) The detected TeV luminosity is of order $L_{\rm TeV} \simeq 3\times 10^{40}$ 
erg/s (if assumed isotropic). It can be shown \cite{rie08} that the test particle 
number density $n_e$, required to account for $L_{\rm TeV}$, does not violate the 
presumed quasi force-free MHD condition in the sense that the corresponding kinetic
energy density $\sim n_e \gamma_2 m_e c^2$ is still much smaller than the allowed 
particle energy density of the background plasma.\\
(3) Is it possible that TeV photons, produced via inverse Compton upscattering, can 
escape from the vicinity of the central black hole? In principle, photons of energy 
$E$ [TeV] will interact most efficiently with target photons in the infrared regime, 
i.e., with energy $\epsilon \simeq (1\,\rm{TeV}/E)$ eV. If the estimated, observed 
infrared luminosity $L_{\rm IR} \sim 10^{41}$ erg/s \cite{why04} is produced on 
characteristic scale $R_{\rm IR}$, the optical depth for $\gamma \gamma$ absorption 
becomes $\tau_{\gamma\gamma}\simeq 0.2\,(L_{\rm IR}/10^{41} \rm{erg/s})
(r_{\rm l}/R_{\rm IR})(E/1\,\rm{TeV})$, indicating that TeV photons can escape even 
if a non-negligible percentage of the observed infrared flux would originate in the
innermost region of the accretion flow $r_l \sim 60\, r_s$. In particular, assuming all of 
the observed flux in M87 to arise in an ADAF (which is perhaps over-restrictive given 
its jet) constraints transparent radii of 10 TeV photons to scales between 5 and 13 
Schwarzschild radii \cite{wan08}, consistent with the scenario presented here. 
\section{Conclusion}
VHE gamma-ray observations of low-luminous, non-blazar AGN jet sources like M87
could provide an ideal test laboratory for particle acceleration close to the 
supermassive black hole horizon and serve as a fundamental diagnostic tool for 
the conditions prevalent in the innermost part of the accretion flow. In blazars, 
these effects are likely to be swamped by relativistically beamed jet emission,
while in luminous quasar sources interactions with the ambient photon field will 
severely limit the efficiency of centrifugal acceleration.\\   
Using a simple toy model, we have shown that efficient centrifugal acceleration of 
electrons in the vicinity of the central supermassive black hole could naturally 
account for the observed VHE characteristics in M87, including variability on time 
scales of about two days and a hard TeV spectrum. The strength of the considered 
scenario is not only that it successfully reproduces the VHE characteristics, but 
also that it allows substantial corroboration of facts by drawing on insight 
gained in accretion disk physics and jet formation theory. 
Generalizing the approach suggests that (i) similar to M87 other low luminous AGN 
jet sources such as, for example Cen~A, may as well be TeV $\gamma$-ray emitting 
sources, and that (ii) efficient centrifugal acceleration may lead to the onset 
of energetic particle beams in BL Lac-type objects, responsible for X-ray and TeV 
radiation on larger ($\gtrsim 10\, r_{\rm L}$) distances \cite{kra07}.

\section*{Acknowledgments}
Support by a LEA Fellowship is gratefully acknowledged.
\balance

\vfill
\end{document}